\documentclass[aps,prl,superscriptaddress,twocolumn,floatfix,10pt]{revtex4-2}

\usepackage{graphicx}
\usepackage[utf8]{inputenc}
\usepackage{amsmath,amsfonts}
\usepackage[colorlinks=true,citecolor=blue]{hyperref}

\begin{document}

\author{Lorenz Meyer}	
\affiliation{Institut für Physik, Technische Universität Ilmenau, D-98693 Ilmenau, Germany}

\author{Jose L.\ Lado}
\affiliation{Department of Applied Physics, Aalto University, 02150 Espoo, Finland}

\author{Nicolas Néel}	
\affiliation{Institut für Physik, Technische Universität Ilmenau, D-98693 Ilmenau, Germany}

\author{Jörg Kröger}
\affiliation{Institut für Physik, Technische Universität Ilmenau, D-98693 Ilmenau, Germany}

\title{Control of Andreev Reflection via a Single-Molecule Orbital}

\begin{abstract}
Charge transport across a single-molecule junction fabricated from a normal-metal tip, a phthalocyanine, and a conventional superconductor in a scanning tunneling microscope is explored as a function of the gradually closed vacuum gap.
The phthalocyanine (2H-Pc) molecule and its pyrrolic-hydrogen-abstracted derivative (Pc) exhibit vastly different behavior.
Andreev reflection across the 2H-Pc contact exhibits a temporary enhancement that diminishes with increasing conductance.
The hybridization of 2H-Pc with the tip at contact formation gives rise to a Kondo-screened molecular magnetic moment.
In contrast, the single-Pc junction lacks Andreev reflection in the same conductance range.
Spectroscopies and supporting nonequilibrium Green function calculations highlight the importance of a molecular orbital close to the Fermi energy for Andreev reflection.
\end{abstract}

\maketitle

Electron transport across a junction with a normal-conductor (NC) and superconductor (SC) electrode is possible for electron energies within the superconducting Bardeen-Cooper-Schrieffer (BCS) energy gap of width $2\Delta$ via Andreev reflection (AR) \cite{spjetp_19_1228} despite the absence of quasielectron states in the SC\@.
To this end, an electron propagating in the NC is retroreflected at the NC--SC interface as a hole in the NC while a Cooper pair is created in the SC\@.
The importance of AR is revealed by, e.\,g., the proximity effect where the superconducting phase is induced in NC materials interfacing an SC on the length scale of the SC coherence length \cite{rmp_36_225}\@.
Moreover, AR can be used as a sensitive probe for spin-polarized currents across point contacts \cite{science_282_85,prl_81_3247}.
The ferromagnet--SC interface has recently attracted substantial interest in the context of AR because of the vividly discussed spin-triplet pairing \cite{prb_89_134517,prl_115_116601}, which may be crucial for superconducting spintronics \cite{natphys_11_307} and topologically protected Majorana bound states \cite{prl_104_040502,prx_6_031016}.
For the new class of magnetic materials, altermagnets \cite{prx_12_031042,prx_12_040501}, AR likewise plays an exceptionally important role \cite{prb_108_054511}.

Charge transport studies of nanoscale junctions with NC and SC leads were experimentally carried out predominantly by quantum dot and point contact measurements \cite{physrep_377_81,worldsci_2017}, and the overwhelming work is of theoretical nature \cite{advphys_60_899}.
The use of a scanning tunneling microscope (STM) that, besides the flexibility of the junction composition, offers the unique capability of characterizing the geometric and electronic structure of the contact is surprisingly scarce.
So far, spectroscopy of AR with an STM has been reported for polycrystalline Pb samples \cite{prb_46_5814}, V$_3$Si(100) \cite{prb_79_144522}, C$_{60}$ on Nb(110) \cite{prl_118_107001}, Mn-phthalocyanine (Mn-Pc) on Pb(111) \cite{prb_97_195429}, and for Pb(110) \cite{prr_3_033248,nl_22_4042}.
For SC tips and SC samples, multiple AR was used to determine charge transport channels for C$_{60}$ molecules on Pb(111) \cite{prb_90_241405} as well as for single-atom junctions on Al(100) \cite{prb_105_165401}, to explore the subharmonic gap structure in asymmetric Nb--Nb contacts \cite{prb_74_132501}, and to unveil the interplay between Yu-Shiba-Rusinov (YSR) states \cite{aps_21_75,ptp_40_435,jetp_29_1101}, Josephson currents \cite{prb_74_132501} and AR \cite{prl_115_087001,prl_121_196803,prb_101_235445,natphys_16_1222}.
Spectroscopy of paramagnetic Mn-Pc on Pb(111) with an SC tip gave rise to a thorough understanding of the competition between Kondo screening and superconducting pair-breaking interactions \cite{science_332_940}.

Here, the behavior of a single phthalocyanine (2H-Pc) and its pyrrolic-H-abstracted derivative (Pc) in an STM NC--SC junction is compared by scanning tunneling spectroscopy (STS) of the differential conductance ($\text{d}I/\text{d}V$, $I$: current, $V$: voltage) \cite{pcar_si}.
The differences are striking.
While for 2H-Pc, AR develops progressively with increasing junction conductance, AR is essentially absent for Pc.
Moreover, AR for 2H-Pc is weakened again with increasing junction conductance, which is at odds with expectations from the Blonder-Tinkham-Klapwijk (BTK) picture \cite{prb_25_4515} of AR\@.
Intriguingly, 2H-Pc -- again strongly contrasting Pc -- shows signatures of a magnetic moment upon hybridizing with the NC tip, which is evidenced by the Kondo effect.
The observations are rationalized by wide-range STS and by Green function simulations, which both indicate the importance of a molecular orbital (MO) close to the Fermi energy ($E_\text{F}$) for the efficiency of AR\@.

\begin{figure}
\centering
\includegraphics[width=\columnwidth]{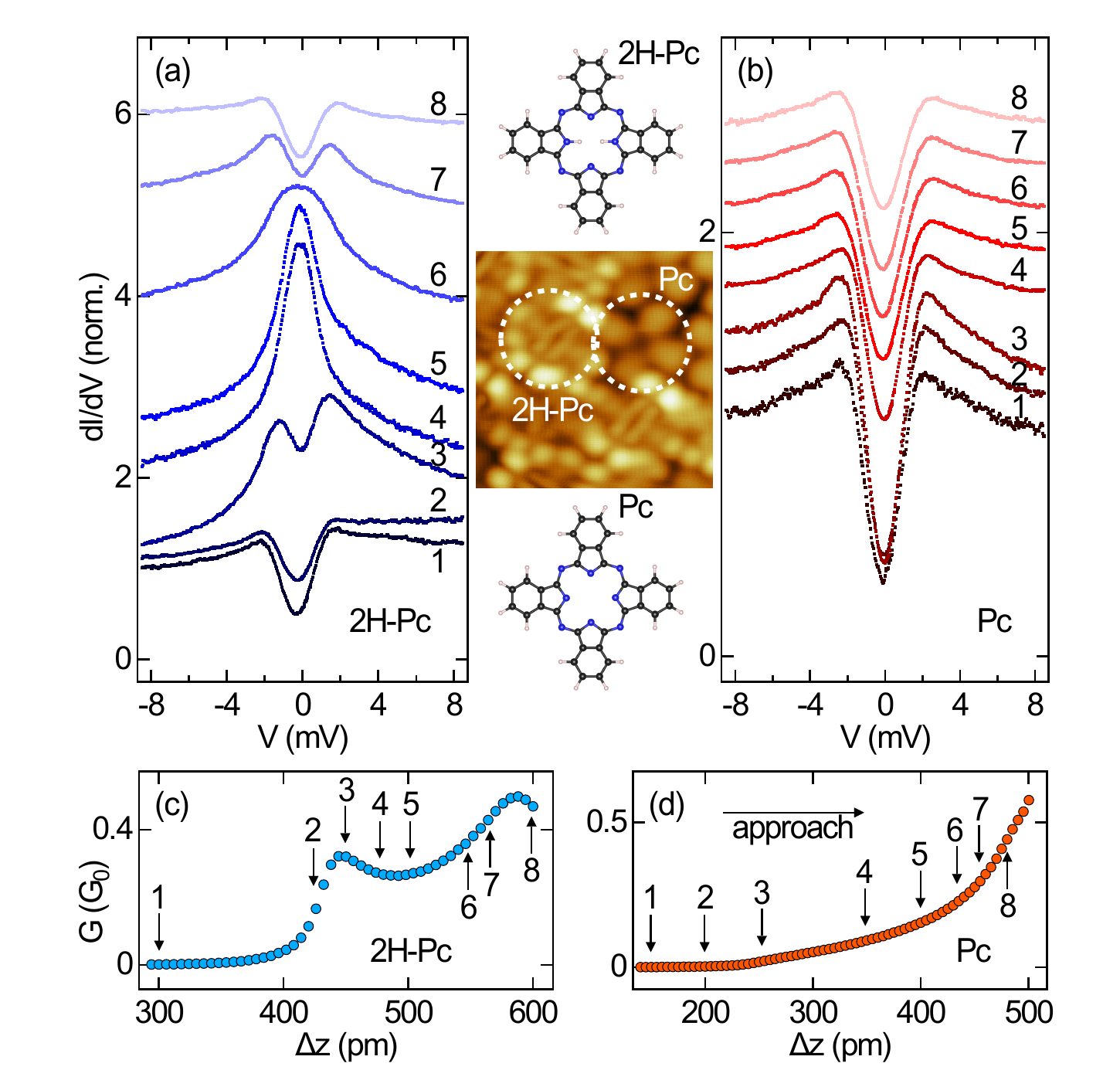}
\caption{(a) Evolution of $\text{d}I/\text{d}V$ spectra acquired atop the 2H-Pc center as a function of $G$ (feedback voltage for all spectra: $200\,\text{mV}$)\@.
Insets: STM image of an assembly of 2H-Pc and Pc molecules ($10\,\text{mV}$, $50\,\text{pA}$, $4\,\text{nm}\times 4\,\text{nm}$) together with ball-and-stick models of their relaxed vacuum structure.
(b) As (a) for the Pc center (feedback voltage for all spectra: $20\,\text{mV}$)\@.
The spectra are normalized to unity at (a) $-20\,\text{mV}$ ($1$--$5$), $20\,\text{mV}$ ($6$--$8$), (b) $20\,\text{mV}$ ($1$--$8$) and are vertically offset.
(c),(d) Junction conductance $G$ in units of the quantum of conductance $\text{G}_0=2\text{e}^2/\text{h}$ (e: elementary charge, h: Planck constant) as a function of tip approach to the center of (c) 2H-Pc and (d) Pc acquired at (c) $200\,\text{mV}$, (d) $20\,\text{mV}$\@.
Tip excursion $\Delta z=0$ is defined by feedback loop parameters of (c) $200\,\text{mV}$, (d) $20\,\text{mV}$ and $50\,\text{pA}$\@.
Arrows indicate $G$ at which STS was carried out in (a) and (b)\@.}
\label{fig1}
\end{figure}

The main experimental results are summarized in Fig.\,\ref{fig1}.
It compares the evolution of the BCS energy gap in STS atop 2H-Pc [Fig.\,\ref{fig1}(a)] and Pc [Fig.\,\ref{fig1}(b)] with decreasing (from bottom to top) separation between an Au-coated W tip and the respective molecule adsorbed on Pb(111)\@.
The variation of the junction conductance $G=I/V$ with the tip displacement $\Delta z$ is depicted in Fig.\,\ref{fig1}(c) for 2H-Pc and in Fig.\,\ref{fig1}(d) for Pc\@.
In the far tunneling range of separations, $\text{d}I/\text{d}V$ spectra exhibit the BCS energy gap for both 2H-Pc and Pc (spectra $1$,$2$)\@.
With increasing $G$ the BCS energy gap measured atop 2H-Pc is gradually filled with spectral weight ($3$) and converted at tip--molecule contact into a peak centered at zero sample voltage ($4$,$5$)\@.
This zero-bias resonance (ZBR) is attenuated again in the contact junction ($6$) and develops a central indentation giving rise to a gaplike spectroscopic structure again ($7$,$8$)\@.
In contrast, for Pc the $\text{d}I/\text{d}V$ spectra remain virtually unaltered in a similar conductance range ($3$--$8$)\@.
These results are independent of the NC tip and were reproduced for $5$ differently prepared tips.
A less exhaustive series of spectra for 2H-Pc and Pc was obtained with the same tip \cite{pcar_si}.
Importantly, the structural and electronic integrity of the junction, which is mandatory for the reliable interpretation of the results, has intermittently been controlled by STM and STS\@.
Probing the intramolecular spatial dependence of the observed effect was hampered by junction instabilities experienced outside the molecule center at elevated currents and sample voltages.
Such spatially resolved STS could in principle unveil the interference of Kondo screening and YSR channels \cite{prl_125_256805} or the entanglement of YSR states due to the intramolecular exchange coupling \cite{prl_126_017001}.

In the following, the origin of the ZBR observed for 2H-Pc and its evolution as a function of $G$ will be clarified.
It is tempting to assign the ZBR to AR alone because it matches the BTK picture of an AR-associated excess current that emerges with improved junction transmission.
However, in additional STS experiments with suppressed superconductivity, a ZBR developed at contact, too \cite{pcar_si}.

\begin{figure}
\centering
\includegraphics[width=\columnwidth]{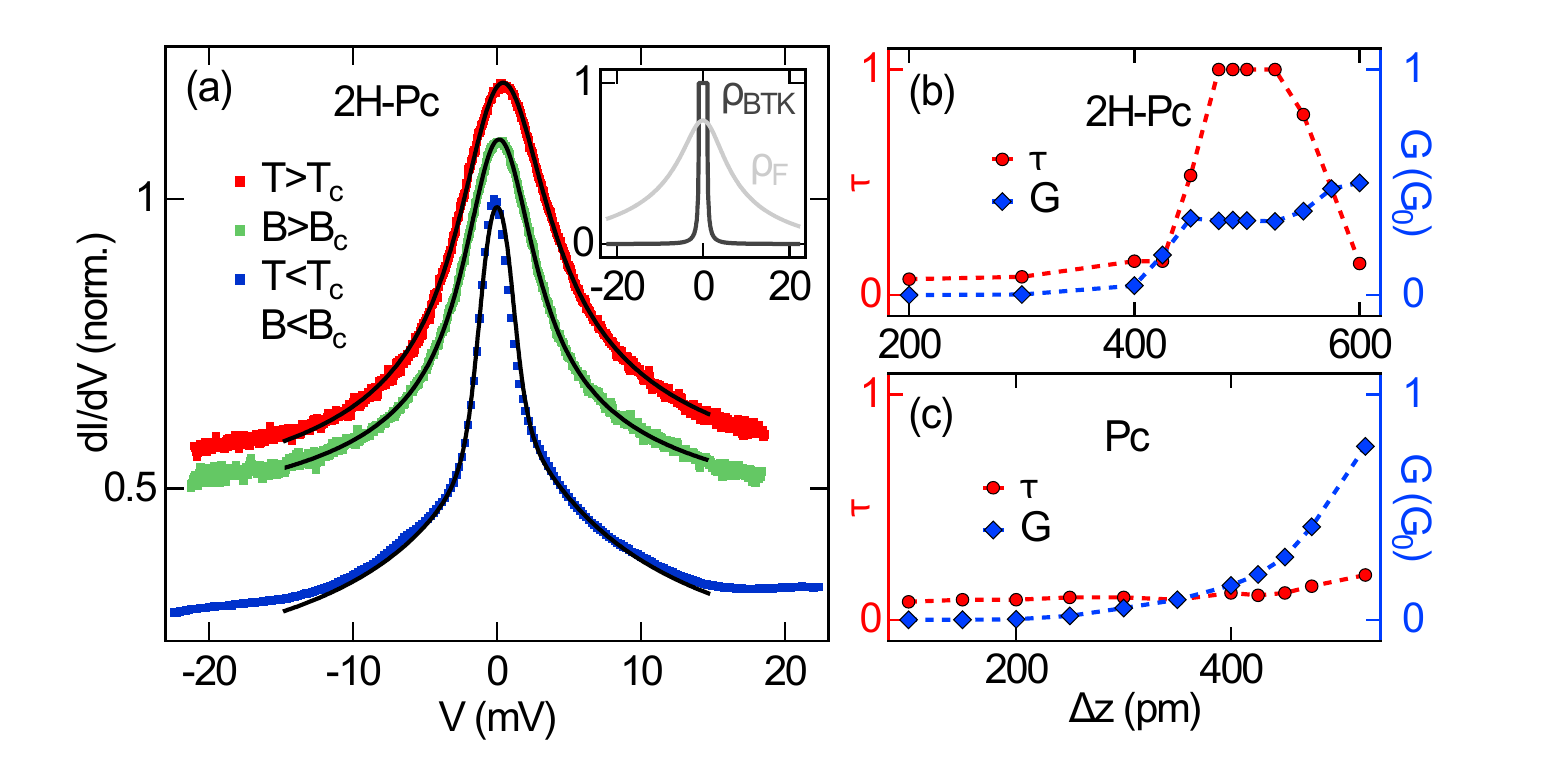}
\caption{(a) Spectra of $\text{d}I/\text{d}V$ acquired atop the 2H-Pc center at $T=5.3\,\text{K}<T_{\text{c}}$, $B<B_{\text{c}}$, $G=0.26\,\text{G}_0$ (bottom), $T<T_\text{c}$, $B=100\,\text{mT}>B_{\text{c}}$, $G=0.24\,\text{G}_0$ (middle), $T=8.0\,\text{K}>T_{\text{c}}$, $B<B_\text{c}$, $G=0.18\,\text{G}_0$ (top) together with fits (solid lines, see text)\@.
Feedback loop parameters: $200\,\text{mV}$ (bottom), $20\,\text{mV}$ (middle, top), $50\,\text{pA}$\@.
Inset: graphs of the BTK ($\varrho_\text{BTK}$) and Frota ($\varrho_\text{F}$) spectral densities used for the fit of the bottom spectrum in (a)\@.
(b) Collection of BTK transmissions $\tau$ and conductances $G$ as a function of tip displacement $\Delta z$ for 2H-Pc.
(c) As (b) for Pc.
Zero tip excursion is defined by feedback loop deactivation at (b) $200\,\text{mV}$, (c) $20\,\text{mV}$ and $50\,\text{pA}$\@.
Dashed lines in (b),(c) are guides to the eye.}
\label{fig2}
\end{figure}

Figure \ref{fig2}(a) shows that for temperatures exceeding the critical temperature $T_\text{c}=7.2\,\text{K}$ \cite{rmp_26_277,rmp_35_1} (top) and for external magnetic fields larger than the critical field $B_\text{c}=80\,\text{mT}$  \cite{rmp_26_277,rmp_35_1} (middle) a broad ZBR appears in $\text{d}I/\text{d}V$ spectra of the 2H-Pc contact junction. 
Its line shape is well described by the Frota profile (solid line) with a full width at half maximum (FWHM) of $\Gamma=7.8\,\text{meV}$, which clearly exceeds the BCS energy gap width of $2\Delta=2.1\,\text{meV}$ (at $4.8\,\text{K}$)\@.
Because of its Frota line shape and its characteristic broadening with increasing temperature and magnetic field \cite{pcar_si} this resonance can reasonably be assigned to the Abrikosov-Suhl resonance (ASR) that signals the Kondo effect \cite{prb_45_1096} with a Kondo temperature of $\approx 41\,\text{K}$ \cite{pcar_si}.  
While free 2H-Pc does not exhibit a magnetic moment \cite{jpcc_124_10441}, its hybridization with a metal surface was previously demonstrated to induce a paramagnetic state and the associated Kondo effect in the metal substrate \cite{jpcc_124_10441,prb_109_L241401}.
In the present case, 2H-Pc hybridizes with the NC tip and adopts a magnetic moment.
The occurrence of YSR states, as seen in $\text{d}I/\text{d}V$ spectroscopy of the 2H-Pc contact junction with an SC tip \cite{pcar_si}, additionally corroborates the induced molecular paramagnetism.
Therefore, the ASR is expected to additionally contribute to the ZBR in 2H-Pc spectra when Pb(111) is in its superconducting state.
Notably, the ASR and the YSR states are only observed for junctions with collapsed tunneling barrier where the tip is in contact with 2H-Pc.
Consequently, the Kondo screening and the YSR states presumably occur in the NC and SC tips, respectively.
While a similar scenario was reported previously for molecules decorating Nb tips \cite{nanoscale_14_15111}, the presence of the Kondo effect and intragap states in the SC sample cannot be excluded.
For Pc, a ZBR is absent in the normal state of Pb(111) and intragap states are missing in the superconducting phase throughout the entire range of tip--molecule distances from tunneling to contact \cite{pcar_si}.

Upon entering the superconducting phase, the ZBR adopts a different line shape [bottom spectrum in Fig.\,\ref{fig2}(a)]\@.
It considerably sharpens and cannot be described by a Frota function alone.
Rather, the resulting line shape including the tails outside the BCS energy gap is best captured by a combination of a Frota and BTK function [inset to Fig.\,\ref{fig2}(a)] \cite{pcar_si}, which evidences the presence of AR in the 2H-Pc junction.
Importantly, matching $\text{d}I/\text{d}V$ spectra with the combined Frota-BTK function for different tip--2H-Pc separations gives rise to the variation of the BTK junction transmission $\tau$ with $\Delta z$ as depicted in Fig.\,\ref{fig2}(b)\@.
It first rises with increasing $\Delta z$ and $G$, reaches $\tau\approx 1$ for $450\,\text{pm}\leq\Delta z\leq 550\,\text{pm}$ and then decreases again.
While the increase of $\tau$ is compatible with the BTK picture of AR owing to a reduced interface barrier, the decrease of $\tau$ is unexpected and must be rationalized (see below)\@.
The essential invariance of $\Gamma$ of the Frota profile reflects an unaffected Kondo effect.
For Pc, in contrast, $\tau$ does not exceed $0.25$ even for a junction conductance close to $\text{G}_0$ [Fig.\,\ref{fig2}(c)]\@.
Consequently, AR and Kondo screening are important for the 2H-Pc junction, while both effects appear to be virtually irrelevant for Pc.

\begin{figure}
\centering
\includegraphics[width=0.8\columnwidth]{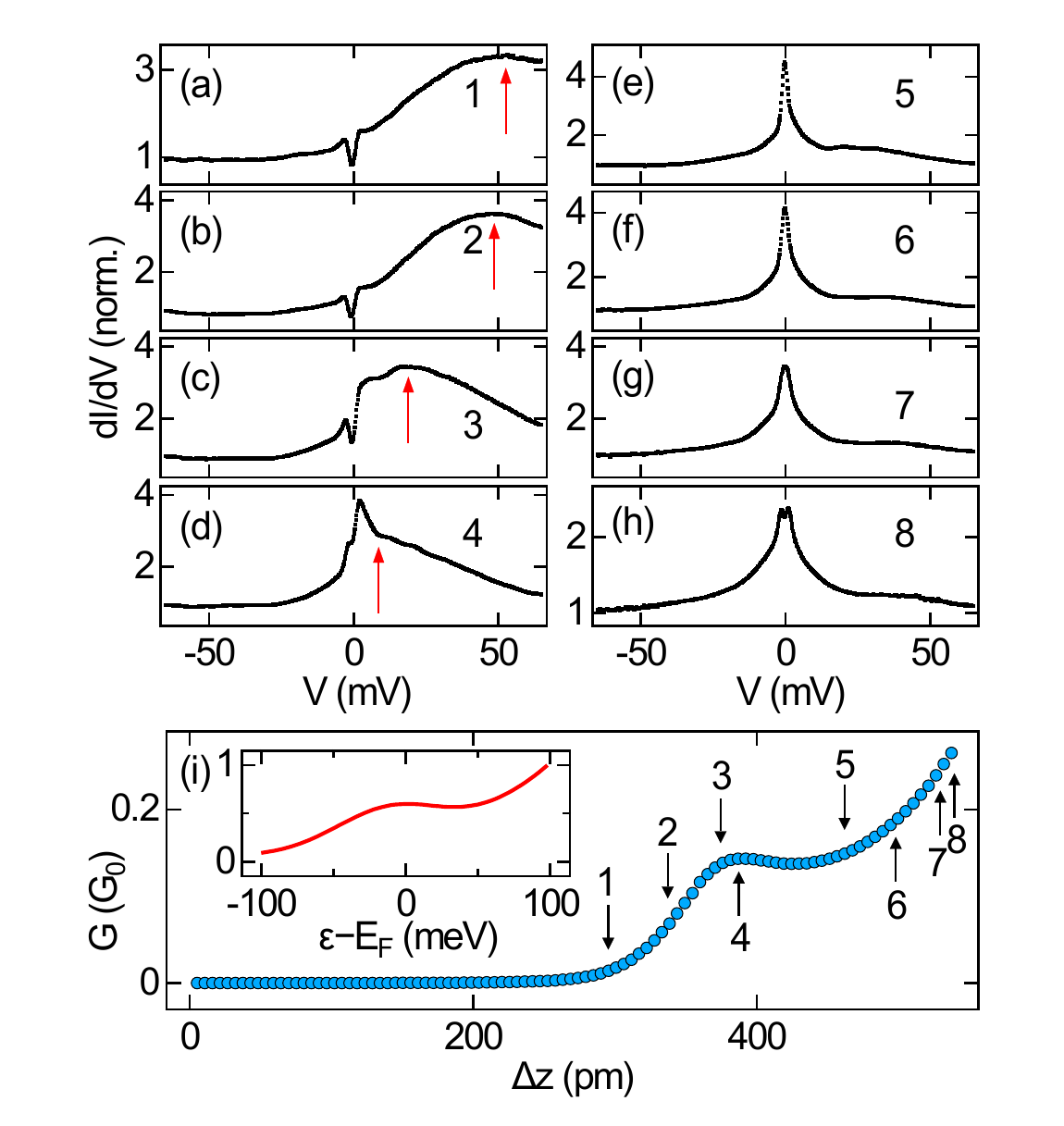}
\caption{(a)--(h) Spectra of $\text{d}I/\text{d}V$ acquired atop the 2H-Pc center with increasing junction conductance ($1$--$8$)\@.
The arrows mark the shifting MO signature.
All spectra are normalized to unity at $-100\,\text{mV}$\@.
(i) $G$-versus-$\Delta z$ trace with marked $G$ for the spectra in (a)--(h)\@.
Inset: simulated variation of $G$ using a Gaussian MO shifting through $E_\text{F}$ (the maximum of $G$ is normalized to unity) \cite{pcar_si}.
Feedback loop parameters for the spectra in (a)--(h) and for defining $\Delta z=0$ in (i): $100\,\text{mV}$, $50\,\text{pA}$\@.}
\label{fig3}
\end{figure}

After clarifying the origin of the low-energy spectroscopic components, the central question as to the occurrence of AR and its unexpected behavior for 2H-Pc as well as the apparent absence of AR in the case of Pc is addressed.
Inspection of the 2H-Pc spectral data [Fig.\,\ref{fig1}(a)] unveils an asymmetry.
For spectra $1$--$5$, the positive-voltage data ride on a higher background than the data at negative voltage, while this asymmetry is reversed for the $\text{d}I/\text{d}V$ data $6$--$8$\@.
In contrast, all spectra of Pc [Fig.\,\ref{fig1}(b)] are symmetric with respect to $V=0$\@.

\begin{figure}
\centering
\includegraphics[width=0.85\columnwidth]{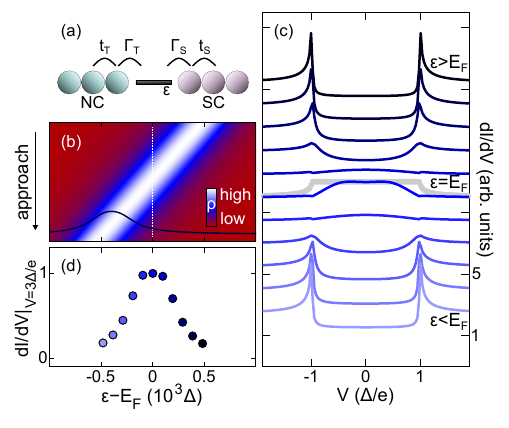}
\caption{(a) Sketch of the one-dimensional junction model underlying the simulations.
The molecule contacted by the NC and SC leads is characterized by a single electron level with variable energy $\varepsilon$.
(b) Collection of calculated MO spectral densities $\varrho$ with decreasing $\varepsilon$ (from top to bottom)\@.
The vertical line marks the position of $E_\text{F}$\@.
(c) Simulated $\text{d}I/\text{d}V$ spectra of the BCS energy gap for varying $\varepsilon$ as shown in (b)\@.
The gray line depicts the BTK function extracted from fits to experimental data \cite{pcar_si}.
(d) Variation of calculated $\text{d}I/\text{d}V$ at $V=3\Delta/\text{e}$ as a function of $\varepsilon$.
For the simulations, $J=100\,\Delta$, $\Gamma_\text{S}=\Gamma_\text{T}=300\,\Delta$, $t_\text{S}=t_\text{T}=1000\,\Delta$ were used.}
\label{fig4}
\end{figure}

To unveil the origin of the spectral background variation, Figs.\,\ref{fig3}(a)--(d) demonstrate by $\text{d}I/\text{d}V$ spectra in a wider $V$ range that the signature of a 2H-Pc MO, presumably of the lowest unoccupied MO \cite{pcar_si}, causes the observed asymmetry.
No such MO is evidenced for Pc \cite{pcar_si}.
With decreasing tip--molecule separation, the 2H-Pc orbital shifts toward $E_\text{F}$ and straddles the Fermi level [arrows in Fig.\,\ref{fig3}(a)--(d)]\@.
The shift of the MO is supposedly induced by the gradual increase of the 2H-Pc hybridization with the tip.
It can readily be tracked until the junction conductance reaches $\approx 0.1\,\text{G}_0$ [Fig.\,\ref{fig3}(d)]\@.
With further increase of $G$, the ZBR becomes dominant [Figs\,\ref{fig3}(e)--(h)] and makes the contribution of the MO to $\text{d}I/\text{d}V$ data difficult to discern.
However, the $G$-versus-$\Delta z$ trace [Fig.\,\ref{fig3}(i)] that underlies the spectra of Figs.\,\ref{fig3}(a)--(h) can be reproduced by assuming in a one-dimensional model of the tunneling barrier \cite{jap_34_1793,jap_34_2581} the MO density of states as a shifting Gaussian with a constant background [inset to Fig.\,\ref{fig3}(i)] \cite{pcar_si}.
Spectrum $5$ with maximal ZBR can therefore be associated with the MO well centered at $V=0$\@.
Consequently, the experimental data strongly hint at an electronic resonance that enhances AR when it is at $E_\text{F}$ while it weakens AR at resonance energies departing from the Fermi level.
In a recent study, the occurrence of Machida-Shibata states in the BCS energy gap of Nb(110) was likewise controlled by the energy of electronic states confined to atomic corrals fabricated on adsorbed Ag islands \cite{nature_621_60}. 

A Green function method for a one-dimensional junction geometry [Fig.\,\ref{fig4}(a)] is used to support the experimental observations.
It assumes a classical spin and, thus, Kondo correlations are out of the scope.
The inclusion of such correlations would likely change the electron transport properties depending on the relevant energy scales, $\text{k}_\text{B}T_\text{K}$ ($\text{k}_\text{B}$: Boltzmann constant, $T_\text{K}$: Kondo temperature) and $\Delta$\@.
Indeed, the ASR was suppressed (present) in quantum dot experiments for $\text{k}_\text{B}T_\text{K}\ll\Delta$ ($\text{k}_\text{B}T_\text{K}\gg\Delta$) \cite{prl_89_256801}, while a scenario of competing Kondo screening and Cooper pairing was reported in an STM study for $\text{k}_\text{B}T_\text{K}\approx\Delta$ \cite{science_332_940}.
Moreover, AR channels were demonstrated to act as Kondo screening channels \cite{prb_81_121308,prb_94_165144}.
Here, the simulations are conceived for exploring the role of an electronic resonance at $E_\text{F}$ on AR\@.

The underlying Hamiltonian $\mathcal{H}=\mathcal{H}_{\text{S}}+\mathcal{H}_{\text{T}}+\mathcal{H}_{\text{M}}+\mathcal{H}_{\text{C}}$ describes the substrate via
\begin{equation}
\mathcal{H}_{\text{S}}=
\mu_S \sum_{i,s} c^\dagger_{i,s}c_{i,s}
+ t_{\text{S}}\sum_{\langle ij\rangle s}c^\dagger_{i,s}c_{j,s}+\Delta\sum_{i}c^\dagger_{i,\uparrow}c^\dagger_{i,\downarrow}+\text{h.c.}    
\label{eq:hs}
\end{equation}
while the NC tip is modeled by
\begin{equation}
\mathcal{H}_{\text{T}} = 
\mu_T \sum_{i,s} d^\dagger_{i,s}d_{i,s} + t_{\text{T}}\sum_{\langle ij \rangle s} d^\dagger_{i,s}d_{j,s}+\text{h.c.}
\label{eq:ht}
\end{equation}

In Eqs.\,(\ref{eq:hs}),(\ref{eq:ht}), $t_{\text{S}}$ and $t_{\text{T}}$ denote the substrate and tip electron
hopping parameters, respectively.
The sums are restricted to nearest neighbors, as indicated by $\langle ij\rangle$\@.
The Hamiltonian of the molecule takes the form
\begin{equation}
\mathcal{H}_{\text{M}}=\varepsilon\sum_{s}f^\dagger_{s}f_{s}+J\sum_{s,s'}\sigma_z^{s,s'}f^\dagger_{s}f_{s'}+\text{h.c.}
\label{eq:hm}
\end{equation}
where $\varepsilon$ and $J$ denote the orbital energy and intramolecular magnetic exchange coupling, respectively. 
The chemical potentials $\mu_\text{S}$ and $\mu_\text{T}$ control the electron-hole asymmetry of the leads (with $\mu_\text{S}=\mu_\text{T}=0$ representing electron-hole symmetry)\@.
The operator $c^\dagger_{i,s}$ ($d^\dagger_{i,s}$) [Eqs.\,(\ref{eq:hs}),(\ref{eq:ht})] creates an electron with spin $s\in\{\uparrow,\downarrow\}$ at position $i$ of the substrate (tip), while $c_{i,s}$ ($d_{i,s}$) annihilates it; for the molecule, these operators are denoted $f^\dagger_s$ and $f_s$ [Eq.\,(\ref{eq:hm})], and $\sigma_z^{s,s'}$ represents the $z$ Pauli matrix. 
The hybridization of the molecular orbital and the two electrodes is modeled as
\begin{equation}
\mathcal{H}_{\text{C}}=\Gamma_{\text{S}}\sum_s f^\dagger_{s}d_{0,s}+\Gamma_{\text{T}}\sum_s c^\dagger_{0,s}f_{s}+\text{h.c.}
\end{equation}
with $\Gamma_{\text{S}}$ and $\Gamma_{\text{T}}$ accounting for the coupling strengths of the molecule to the substrate and the tip, respectively. 

An important ingredient of the model is the adjustable single molecular energy level $\varepsilon$.
Figure \ref{fig4}(b) shows the calculated MO spectral density shifting from $\varepsilon=0.7\,t_\text{S}$ above (top) to $\varepsilon=-0.7\,t_\text{S}$ below (bottom) $E_\text{F}$\@.
The spectral density is computed as $\varrho(\varepsilon)=-\Im\{\text{Tr}[P_e(\varepsilon-\mathcal{H}_\text{M}-\Sigma_\text{S}-\Sigma_\text{T})^{-1}]\}/\pi$, where $\Sigma_{\text{S,T}}=|\Gamma_{\text{S,T}}|^2 G^\text{surf}_{\text{S,T}}$ are the self-energies of substrate and tip, $G^\text{surf}_{\text{S,T}}$ the surface Green functions, and $P_e$ the projection operator in the particle Nambu sector \cite{cup_1995}.
The associated evolution of the simulated BCS energy gap is depicted in Fig.\,\ref{fig4}(c)\@.
With decreasing $\varepsilon$ (top to bottom), the intragap region is filled with spectral weight due to AR\@.
Maximum AR is reached when $\varepsilon=E_\text{F}$\@.
The AR signal is then attenuated again upon further decreasing $\varepsilon$ with an eventually recovered BCS energy gap.
To see the dependence of AR on the MO more clearly, Fig.\,\ref{fig4}(d) presents the variation of the calculated $\text{d}I/\text{d}V$ data at $V=3\Delta/\text{e}$.
It was previously shown that the BTK transmission scales with the normal-state $\text{d}I/\text{d}V$ \cite{prl_118_107001}, which is well reached at this sample voltage.
Indeed, the evolution of $\left.\text{d}I/\text{d}V\right|_{V=3\Delta/\text{e}}$ is similar to the observed trend of $\tau$ in the experiments [Fig.\,\ref{fig2}(b)]\@.
In a previous report on single-Al contacts with SC electrodes (Al tip, Al sample), a reduction of the transmission of electron transport channels was inferred from the excess current due to multiple AR and tentatively traced to relaxations of the junction \cite{prb_105_165401}.
Before concluding, it is noteworthy that an increased $J$ in the simulations leads to an exchange-split MO, the concomitant partial suppression of AR \cite{science_282_85,prl_81_3247}, and the occurrence of YSR states \cite{pcar_si}.

In conclusion, the efficiency of AR across a single-molecule contact can be tuned by the energy of an electronic resonance close to the Fermi level.
This finding offers the unique opportunity to control the proximity effect at the NC--SC interface, which is particularly appealing at interfaces involving magnetic materials owing to their potential in the observation of unconventional pairing and the construction of the Majorana bound state.
Moreover, the results offer a strategy to identify zero-bias anomalies, which has become decisive in spectroscopies of low-energy excitations in superconducting junctions \cite{natrevmater_3_52}. 

\acknowledgments{Funding by the Deutsche Forschungsgemeinschaft (Grant Nos.\ KR 2912/18-1 and KR 2912/21-1) and the German Federal Ministry of Education and Research within the ''Forschungslabore Mikroelektronik Deutschland (ForLab)'' initiative as well as discussions with Maximilian Kögler are acknowledged.
JLL acknowledges the computational resources provided by the Aalto Science-IT project, and the support from the Research Council of Finland (Grant Nos.\ 331342 and 358088)\@.}

\textit{Data availability}---The supporting data for this article are openly available at Zenodo \cite{pcar_zenodo}.

%

\end{document}